\documentclass[twocolumn,showpacs,preprintnumbers,amsmath,amssymb]{revtex4}
\usepackage{graphicx}% Include figure files
\usepackage{dcolumn}% Align table columns on decimal point
\usepackage{bm}% bold math

\begin{document}

%\preprint{}

\title{Light by light diffraction in vacuum}

\author{Daniele Tommasini and Humberto Michinel}

\affiliation{Departamento de F\'\i sica Aplicada, Universidade de Vigo,
As Lagoas, s/n. E-32004 Ourense, Spain.}

\date{March 24, 2010}

\begin{abstract}
We show that a laser beam can be diffracted by a
more concentrated light pulse due to quantum vacuum effects.
We compute analytically the intensity pattern in a 
realistic experimental configuration, 
and discuss how it can be used to
measure for the first time 
the parameters describing photon-photon scattering in vacuum.
In particular, we show that the Quantum Electrodynamics prediction
can be detected in a single-shot experiment 
at future 100 petawatt lasers such as ELI or HIPER.
On the other hand, if carried out at one of the
present high power facilities, such as OMEGA EP, 
this proposal can lead 
either to the discovery of non-standard physics,
or to substantially improve the current PVLAS 
limits on the photon-photon cross section 
at optical wavelengths.
This new example of manipulation of light by light 
is simpler to realize and more sensitive than existing, alternative proposals, 
and can also be used to test Born-Infeld theory or to search for 
axion-like or minicharged particles.
\end{abstract}

%\pacs{12.20.Ds, 42.50.Xa, 12.60.-i, 42.25.Fx}

\maketitle

The linear propagation of light in vacuum, as described by Maxwell equations,
is a basic assumption underlying our communication system, allowing e.g.~that 
different electromagnetic waves do not keep memory of their possible crossing
in the way to their reception points. However, 
this superposition principle is expected to be violated
by quantum effects. In fact, Quantum Electrodynamics (QED)
predicts the existence of Photon-Photon Scattering 
in Vacuum (PPSV) mediated by virtual charged particles 
running in loop diagrams~\cite{Costantini1971},
although the rate is negligible in all the experiments that have
been performed up to now. On the other hand, additional, possibly larger contributions 
to the process may appear in non-standard models such as Born Infeld 
theory~\cite{BornInfeld,Denisov,Gaete}
or in new physics scenarios involving minicharged~\cite{MCP} 
or axion-like~\cite{ALP} particles. Therefore, the search for PPSV
is important not only to demonstrate a still unconfirmed,
fundamental quantum property of light,
but also to either discover or constrain these kinds of new physics.

In the last few years, there has been an increasing interest in 
the quest for PPSV~\cite{PPSsearch,PVLAS,JHEP09,NaturePhotonics2010}.
Here, we present a new scenario to search for this phenomenon
using ultra-high power lasers~\cite{mourou06}.
In our proposal, two almost contra-propagating 
laser pulses cross each other. 
Due to PPSV, the more concentrated pulse behaves like a phase object
diffracting the wider beam. 
The resulting intensity pattern can then be observed on a screen, 
and will correspond to a direct detection of 
scattered photons.

{\em The effective Lagrangian for photon-photon scattering.} 
Following Ref.~\cite{JHEP09}, we will assume that
for optical wavelengths the electromagnetic fields ${\bf E }$ and ${\bf B}$ are
described by an effective Lagrangian of the form
\begin{equation}
{\cal L}={\cal L}_0
+\xi_L {\cal L}_0^2+ \frac{7}{4}\xi_T {\cal G}^2,
\label{L}
\end{equation}
being ${\cal L}_0=\frac{\epsilon_0}{2}\left({\bf E}^2-{c^2\bf B}^2\right)$
the Lagrangian density of the linear theory and 
${\cal G}=\epsilon_0 c({\bf E}\cdot {\bf B})$. 

In Quantum Electrodynamics, ${\cal L}$ would be 
the Euler-Heisenberg effective Lagrangian~\cite{Euler-Heisenberg}, 
that coincides with Eq.~(\ref{L}) with the identification
$\xi_L^{QED}=\xi_T^{QED}\equiv \xi$, being
\begin{equation}
\xi=\frac{8 \alpha^2 \hbar^3}{45 m_e^4 c^5}\simeq 6.7\times
10^{-30}\frac{m^3}{J}.
\label{constant_xi}
\end{equation}

However, in Born-Infeld theory~\cite{BornInfeld,Denisov,Gaete}, 
or in models involving a new minicharged (or milli-charged)~\cite{MCP}
or axion-like~\cite{ALP} particle, $\xi_L$ and $\xi_T$ will have different
values, as computed in Ref.~\cite{JHEP09}. 

On the other hand, the current 95\% C.L. limit on PPSV at optical wavelengths
has been obtained by the PVLAS collaboration~\cite{PVLAS}. 
As shown in Ref.~\cite{JHEP09}, 
it can be written as 
\begin{equation}
 \frac{\vert 7\xi_T-4\xi_L\vert}{3}<3.2\times10^{-26}\frac{m^3}{ J}.
\label{PVLAS}
\end{equation}
Assuming $\xi_L=\xi_T\equiv\xi^{exp}$ as in QED, 
this can be translated in the limit
$\xi^{exp}<3.2 \times 10^{-26}{m^3}/{J}$, which is
$4.6\times 10^{3}$ times higher than the QED value of Eq.~(\ref{constant_xi}).

\begin{figure}[htb]
{\centering \resizebox*{1\columnwidth}{!}{\includegraphics{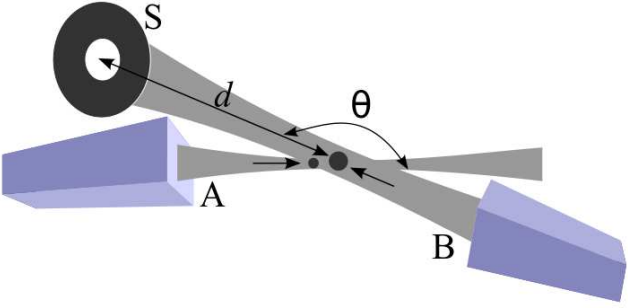}} \par}
\caption{Sketch of the proposed experiment. 
An ultra-intense laser pulse A and a wider probe beam B, both moving 
in a high vacuum, are focused
to a region where they collide at an angle $\theta$ close to $\pi$. 
The diffracted part of the probe is then observed at a distance
$d$ on the ring screen S. 
(In a minimal version, a single laser can produce both beams.)
}
\label{fig1}
\end{figure}

{\em Proposal of an experiment: analytical computations.} 
In our present proposal, illustrated in Fig.~\ref{fig1}, 
a polarized ultrahigh power Gaussian pulse A of transverse width $w_A$ 
crosses an almost contra-propagating polarized `probe' laser pulse B of width $w_B \gg w_A$.
For simplicity, we assume that the two beams have the same 
mean wavelength $\lambda=2 \pi/k$ and frequency $\nu=c/\lambda=c k/2\pi$, 
although in principle they may have different durations $\tau_A$ and $\tau_B$.
We also suppose that the uncertainty in frequency $\Delta \nu$ is much smaller than $\nu$, 
in such a way that we can consider the pulses as being monochromatic with a good approximation. 
Similarly, we assume that the uncertainty in the components of the 
wave vector are much smaller than $k$. 

From Ref.~\cite{JHEP09}, we learn that
the central part of the probe B, 
after crossing the pulse A, acquires a phase shift
\begin{equation}
\phi_{L,T}(0)= I_A(0) k \tau_A a_{L,T}\xi_{L,T}
\label{phase_shift_0} 
\end{equation}
where $I_A(0)$ is the peak intensity of the high power beam at the crossing point,
the indexes $L$ and $T$ refer to the two beams having 
parallel or orthogonal linear polarizations, respectively, 
and we have defined $a_L=4$ and $a_T=7$.

Let $ A_A=A_{A}(0)\exp\left(-\frac{r^2}{w_A^2}\right)$ 
and $A_B=A_{B}(0)\exp\left(-\frac{r^2}{w_B^2}\right)$
describe the dependences of the non-vanishing components of 
the two waves on the radial coordinate $r\equiv\sqrt{x^2+y^2}$ orthogonal 
to the direction of the motion, chosen in the $z$-axis.
The intensity of the pulse A in the colliding region 
will then have the transverse distribution  
$I_A=I_{A}(0)\exp\left(-2\frac{r^2}{w_A^2}\right)$.
As a consequence, the space-dependent phase shift of the wave B just after the collision
with the beam A is
\begin{equation}
\phi(r)=\phi(0)\exp\left(-\frac{2 r^2}{w_A^2}\right),
\end{equation}
where we understand one of the sub-indexes $L$ or $T$.

%%%%%%%%%%%%%%%%%%%%%% FIGURE 1 %%%%%%%%%%%%%%%%%%%%%%%
%%%%%%%%%%%%%%%%%%%%%%%%%%%%%%%%%%%%%%%%%%%%%%%%%%%%%%%%%

Due to this phase shift, the shape of the pulse B 
becomes $A_B=A_{B}(0)\exp\left[-\frac{r^2}{w_B^2}+i\phi(r)\right]$.
As discussed in Ref.~\cite{JHEP09}, $\phi$ is expected to be very small 
at all the facilities that will be available in the near future. 
Therefore
$\exp[i\phi(r)]\simeq 1+i\phi(r)$ 
with a very good approximation, 
and we obtain
\begin{equation}
A_B= A_{B}(0)\left[\exp\left(-\frac{r^2}{w_B^2}\right)
+i\phi(0) \exp\left(-\frac{r^2}{w_0^2}\right)\right],
\label{A_B_after}
\end{equation}
where we have defined $w_0\equiv (2/w_A^2+1/w_B^2)^{-1/2}$.

After the collision, the field $A_B$ propagates linearly,
so that we can just sum the free evolution of each term in Eq.~\eqref{A_B_after},
that can be computed with the approximation
$\omega= c \sqrt{k^2+k_\perp^2}\simeq c( k+k_\perp^2/2 k)$
for the angular frequency, 
where ${\bf k}_\perp=(k_x,k_y,0)$, assuming that 
$\Delta k_{x,y}=1/w\ll k$. 
As a result, the linear evolution of $A_B$ 
on the screen-detector plane $z=d$ produces an intensity pattern 
$I(r)= I_U(r)+I_D(r)+I_I(r)$,
where $I_U$ and $I_D$ correspond to the undiffracted and diffracted waves
respectively,
and $I_I$ represents the interference term. 
The result is 
\begin{eqnarray}
 &I_U(r)=I_B(0) \frac{w_B^2}{w_U^2}\exp\left(-\frac{2 r^2}{w_U^2}\right)
 \nonumber,&\\ 
 &I_D(r)=I_B(0) \phi(0)^2\frac{w_0^2}{w_D^2}\exp\left(-\frac{2 r^2}{w_D^2}\right), 
\label{intensities}
\end{eqnarray}
where we have defined the widths of the undiffracted and diffracted patterns,
$w_U\equiv w_B\sqrt{1+(2 d/k w_B^2)^2}$ and $w_D\equiv w_0\sqrt{1+(2 d/k w_0^2)^2}$, 
and $I_B(0)$ is the peak intensity of the wave B 
at the collision point, that can be related to 
the total power $P_B=P_U$ of the pulse B as $P_U=\frac{\pi}{2} w_B^2 I_B(0)$.
The interference term $I_I(r)$ can be evaluated by multiplying 
$2 \sqrt{I_U(r) I_D(r)}$ by the factor
\begin{equation}
 \sin\left[\frac{d \lambda (w_0^4-w_B^4) r^2}{\pi w_B^2 w_0^2 w_U^2 w_D^2}+\arctan(\frac{d \lambda}{\pi w_0^2})
-\arctan(\frac{d \lambda}{\pi w_B^2})\right], 
\nonumber
\end{equation}
and turns out to be numerically negligible, 
as compared to $I_U(r)+I_D(r)$, in all the configurations that
we will discuss below.

On the other hand, the total power of the diffracted pulse can be obtained by 
integrating Eq.~\eqref{intensities} in the screen plane, 
so that $P_D=\frac{\pi}{2} w_0^2 \phi(0)^2 I_B(0)$,
which is much smaller than $P_U$. 
However, an interesting feature of Eq.~\eqref{intensities} is that 
the diffracted wave is distributed in an area of width 
$w_D\sim 2 \sqrt{2} d/ k w_A\gg w_U\sim 2 d/ k w_B $
(for $2 d\gg k w_B^2$), so that 
it can be separated from the undiffracted wave e.g.~by making a hole in the screen. 
Let $r_0$ be the radius of the 
central region that is eliminated from the screen. We require that the total power 
$P_D(r>r_0)=\frac{\pi}{2} w_0^2 \phi(0)^2 I_B(0)\exp(-2 r_0^2/w_D^2)$
due to the diffracted wave for $r> r_0$ is much larger than the the power of 
the undiffracted wave in the same region, 
$P_U(r>r_0)=\frac{\pi}{2} w_B^2 I_B(0)\exp(-2 r_0^2/w_U^2)$.
A safe choice can be $P_D(r>r_0)=100 P_U(r>r_0)$, that implies
\begin{equation}
r_0=w_D w_U \sqrt{\frac{\log\left(\frac{10 w_B}{\phi(0) w_0}\right)}
{w_D^2-w_U^2}}.
\end{equation}

Finally, using Eqs.~\eqref{phase_shift_0} and \eqref{intensities},
we can compute the number of diffracted photons that will 
be detected after ${\cal N}$ repetitions of the experiment in 
the ring region $r_0<r<R$ of the screen, being $R$ its external radius.
We obtain 
\begin{equation}
 N_D^{\cal N}=\frac{8 f {\cal N}}{\pi \hbar c}\frac{E_A^2 E_B w_0^2}{\lambda w_A^4 w_B^2 }
\left(e^{-\frac{2 r_0^2}{w_D^2}}-e^{-\frac{2 R^2}{w_D^2}}\right) 
(a\xi)_{L,T}^2,
\label{N_D}
\end{equation}
where $f$ is the efficiency of the detector,
and $E_A=P_A\tau_A$ and $E_B=P_B\tau_B$ are the total energies 
of the two pulses. 

{\em Angular constraints and Optimization of the sensitivity.} 
Eq. \eqref{N_D} shows that the number of scattered photons
is proportional to the product $E_A^2 E_B$ of the energies of the two laser beams.
It would then be convenient to use a ultrahigh power pulse also for the 
probe B. This can be done economically by producing both beams simultaneously,
e.g.~by dividing a single pulse of energy $E=E_A+E_B$ before the last focalizations.
The maximum value for $N_D$
is then obtained by taking $E_A=2 E/3$ and $E_B=E/3$. 

The other parameters that can be adjusted in order to maximize $N_D$ are 
the widths $w_A$ and $w_B$ of the two colliding beams. 
The choice of $w_A$ is constrained by the requirement that
the pulse A must not spread
in a significant way during the crossing,
so that $w_A\gtrsim \sqrt{c\tau_B\lambda/\pi}$.
However, a more stringent constraint, involving also the angle 
$\theta$, originates from the condition that the center of pulse A 
has to remain close to the central part of beam B during the interaction.
This implies that $c \tau_B \tan(\pi-\theta)\ll w_A$. A safe 
choice can then be $c \tau_B \tan(\pi-\theta)=w_A/10$. 
On the other hand, the angle 
$\pi-\theta$ has to be such that, out of the collision point, 
the trajectories of
the two beams are separated by a distance sufficiently larger than 
their width.
We conservatively ask that such a distance is 6 times 
the width $\sim z \lambda/\pi w_A$ of the beam A 
at the distance $z$, although one can keep in mind that 
smaller separations, if they turned out to be experimentally 
viable, would allow for better sensitivities. 
For small $\pi-\theta$, we then have 
$\pi - \theta\simeq 6 \lambda/\pi w_A$,
and we can solve for $w_A$, 
\begin{equation}
 w_A=\sqrt{60 c \tau_B\lambda/\pi}.
\label{w_A}
\end{equation}
On the other hand, the value of $w_B>w_A$ that maximizes $N_D$ 
will be computed numerically, and the outer radius $R$ will be chosen 
slightly larger than $\sqrt{2}w_D\sim 2 \lambda d/\pi w_A$, by requiring that 
only a few percent of the diffracted wave is lost. 

Finally, the measurement of the number of diffracted photons 
$N_D$ can be used to determine the values of the parameters 
$\xi_L$ and $\xi_T$.
To evaluate the best possible sensitivity, we will suppose that 
the background of thermal photons and the dark count of the detectors
can be made much smaller than the signal. Although this goal may be difficult 
in practice, in principle it can be achieved
by cooling the ring detector and covering it with a filter that selects a tiny window
of wavelengths around $\lambda$, and
by optically isolating the experimental area 
within the time of response of the detector, 
which should be as small as possible (the ultrashort time 
of propagation of the beams can be neglected in comparison).
Of course, the actual background should be measured by performing
the control experiments in the absence of the beams, and with only one
pulse at a time. 

Under these assumptions, the best sensitivity would correspond to the detection
of, say, 10 diffracted photons, so that the zero result could be excluded 
within three standard deviations.
The ideal, minimum values of $\xi_L$ and $\xi_T$ 
that could be measured would then be
given by Eq.~\eqref{N_D}, 
taking $N_D^{\cal N}=10$ and all the choices reviewed above.
(In the numerical computations that we present below, we also include a
small correction $\sin^4(\theta/2)$ that appears in the expression of 
$\phi(0)$ as shown in Ref.~\cite{JHEP09}.)

{\em Sensitivity at future 100 Petawatt laser facilities.}  
Let us now study the possibility of performing
our proposed experiment, in its economical version discussed above, 
using a 100 Petawatt laser such as ELI~\cite{ELI}
or HIPER~\cite{HIPER}, that are expected to become operative in few years.
In this case, we can use the following values:
total power $P=10^{17} W$, duration $\tau=30 fs$, energy $E=3 kJ$ and wavelength
$\lambda=800 nm$.
With these data, using Eq.~\eqref{w_A}, we can compute 
the suggested value $w_A=12\mu m$ for the width of the spot to which the pulse A 
should be focalized at the collision point. 
Taking e.g.~$d=1 m \gg w_B^2/\pi\lambda$,
we find numerically that the best choice for $w_B$ is $w_B\simeq 5w_A=59\mu m$. 
We then obtain the value $r_0=2.1 cm$ for the central hole in the screen, 
$R=4.8 cm$ for its outer radius, and $w_U=0.43 cm$ and $w_D=3.1 cm$ for the 
widths of undiffracted and diffracted waves. 
The focused intensities of the two beams are
$I_A(0)=3.1\times10^{22}W/cm^2$ and $I_B(0)=6.2\times10^{20}W/cm^2$, and
the angle $\pi-\theta=0.13 rad$. Even if we assume an efficiency as small as 
$f=0.5$, which is a realistic value today for $\lambda\sim800nm$, 
we obtain that our proposed experiment can resolve $\xi_L$ and $\xi_T$ 
as small as $\xi_L^{\rm limit}=2.8\times 10^{-30}{\cal N}^{-1/2}m^3/J$ and
$\xi_T^{\rm limit}=1.6\times 10^{-30}{\cal N}^{-1/2}m^3/J $, values that
are well below the QED prediction even for ${\cal N}=1$. 
Therefore, ELI and HIPER will be able to detect PPSV
at the level predicted by QED in a single shot
experiment. Two single-shot experiments, 
using parallel and orthogonal polarizations 
of the colliding waves respectively, 
would allow to measure both $\xi_L$ and $\xi_T$.

We can restate this result in terms of the number of 
diffracted photons per pulse that will be be scattered in the ring detector.
In the experiment with orthogonal polarizations, this number is $\sim 340$, 
which is two orders of magnitude higher that the value obtained in 
the scenario of Ref.~\cite{NaturePhotonics2010}. 
This indicates that our present proposal is by far more sensitive 
than that of Ref.~\cite{NaturePhotonics2010},
in spite of the fact that we have applied much more realistic assumptions on 
the width $w_A$ of the higher power pulse in order to clearly separate the beams
out of the crossing region. Moreover, 
the configuration of Ref.~\cite{NaturePhotonics2010}
is less economical, since it requires an additional high power laser, 
and it will also present a greater experimental difficulty, 
as far as it needs the alignment of three ultrashort pulses 
(two of them having a spot radius as small as $0.8\mu m\sim\lambda$).

Let us now compare our present proposal with that of Ref.~\cite{JHEP09}. 
The main differences between the two scenarios, both based in the crossing of 
two contra-propagating beams, are the following: 1) in Ref.~\cite{JHEP09}, 
the probe pulse could have much smaller power;
2) in Ref.~\cite{JHEP09}, the two beams had the same width, and the effect of 
PPSV was observed by measuring the phase shift of the probe 
by comparing it with a third beam that was originally in phase with it. 
Actually, the sensitivity calculated in Ref.~\cite{JHEP09} corresponded to 
a greater power, $P\sim 10^{18} W$, and shorter $\tau\sim 10^{-14} s$, 
giving $w\sim 3\mu m$ for $I\sim 10^{25}W cm^{-2}$.
To compare that result with the present proposal, we 
have to use the same power, $P\sim 10^{17} W$, and restate the
sensitivity obtained in a single shot in Ref.~\cite{JHEP09} 
as $\xi_L^{\rm lim}=2.8\times 10^{-8}w^2\lambda/ (16 E)$
and $\xi_T^{\rm lim}=2.8\times 10^{-8}w^2\lambda/ (28 E)$,
where $E=P\tau$ and $w$ are the total energy and focused width 
of the high power beam.
It can be seen that a safe choice for the angle $\theta$ 
in that configuration would have needed 
$w\simeq c\tau$, so that even at $P=10^{17}W$
the setup of Ref.~\cite{JHEP09} would have preferred to use 
$\tau\sim10fs$ instead of $30fs$.
This would lead to $\xi_L^{\rm lim}\simeq 1.3\times 10^{-29}m^3/J$
and $\xi_T^{\rm lim}\simeq 7.2\times 10^{-30}m^3/J$. 
These results would be significantly worse than those that we have obtained
above in the present proposal. Moreover, an additional advantage of our new 
configuration
is the fact that it can be systematically improved by increasing 
the number ${\cal N}$ of crossing events in Eq.~\eqref{N_D}.

{\em Sensitivity at present Petawatt lasers.}  
Facilities such as OMEGA EP~\cite{OMEGA}
can already provide $P=10^{15} W$, $\tau=1ps$ and $E=1kJ$ at $\lambda=1053 nm$.
From Eq.~\eqref{w_A}, we obtain $w_A=78\mu m$. Taking $d=1 m$,
we find numerically the optimal choice $w_B= 5 w_A=0.39mm$,
leading to $r_0=5.2mm$, $R=10 mm$, $w_U=0.95 mm$, $w_D=6.2 mm$, 
$I_A(0)=7.0\times10^{18}W/cm^2$, $I_B(0)=1.4\times10^{17}W/cm^2$ and
$\pi-\theta=0.026rad$. Unfortunately, at present the efficiency 
of photon detectors for $\lambda\sim1\mu m$ is just $f\simeq 0.1$. 
Nevertheless, we obtain that our proposed experiment can resolve $\xi_L$ and $\xi_T$ 
as small as $\xi_L^{\rm limit}=2.1\times 10^{-27}{\cal N}^{-1/2}m^3/J$ and
$\xi_T^{\rm limit}=1.2\times 10^{-27}{\cal N}^{-1/2}m^3/J$. 
Even for the single shot experiment, ${\cal N}=1$, this result is 
at least an order of magnitude below the
current limit of Eq.~\eqref{PVLAS}. As a result, this experiment
at present facilities can already 
either detect PPSV of non-standard origin,
or substantially improve the limits on 
$\xi_L$ and $\xi_T$. 
In the former case, 
the measurement of both $\xi_L$ and $\xi_T$ can be used to
discriminate between different kind of new physics, such as Born Infeld
theory~\cite{BornInfeld,Denisov,Gaete} or models involving 
minicharged~\cite{MCP} or axion-like~\cite{ALP} particles,
using the expressions for the corresponding contributions 
that have been computed in Ref.~\cite{JHEP09}.

Finally, we note that in the minimal realization of 
our proposal, using only one high energy laser pulse
divided in two parts of the same time duration $\tau_A=\tau_B=\tau$, 
the optimal $w_B$ turns out to be proportional to $ w_A$ 
(usually by a factor $\sim 5$).
Taking into account Eqs.~\eqref{N_D} and \eqref{w_A}, this implies that 
the discovery potential $ N_D^{\cal N}$ for PPSV is proportional to
$f P^3 \tau/\lambda^3$. One could then study the possibility
of using a $10PW$ laser having a wavelength in the visible window,
in order to compensate in part the lower power (as compared with the
$100 PW$ case discussed above) with the higher $f$ and $\lambda^{-3 }$ factors. 
In this case, assuming $\tau\sim30fm$,
our computations show that PPSV at the QED rate can be detected 
by accumulating a number of repetitions ${\cal N}\sim 10$. 

We thank D. Novoa and F. Tommasini for useful discussions and help.
This work was supported by the Government of Spain
(project FIS2008-01001) and the University of Vigo (project 08VIA09).

\end{document}